\documentstyle[aps,twocolumn]{revtex}

\input epsf
\tightenlines
\begin{document}
\draft
\preprint{VECC-2000-}
\title{
Scaling of hadronic transverse momenta in a hydrodynamic treatment of
relativistic heavy ion collisions}
\author{Dinesh Kumar Srivastava}
\address{Variable Energy Cyclotron Centre, 1/AF Bidhan Nagar, Kolkata
700 064, India}
\date{\today}
\maketitle

\begin{abstract}

The transverse momenta of hadrons in central nucleus-nucleus collisions
are evaluated in a boost invariant hydrodynamics with transverse
expansion. Quark gluon plasma is assumed to be formed in the initial state
which expands and cools via a first order phase transition to a
rich hadronic matter and ultimately undergoes a freeze-out.
The average transverse momentum of pions, kaons, and protons is
estimated for a wide range of multiplicity densities and transverse
sizes of the system. For a given system  it is found to scale 
 with the square-root of the particle rapidity 
density per unit transverse area, and consistent with the corresponding
values seen in $p\overline{p}$ experiments at 1800 GeV, suggesting a
universal behaviour.  The average
transverse momentum shows only an approximate scaling with multiplicity
density per nucleon which is at variance with the $p\overline{p}$
data.  

\end{abstract}
\narrowtext
 
The last two decades have witnessed an active pursuit of relativistic heavy
ion collisions in order to create a quark-gluon plasma,
first `seen' a few micro-seconds after the big-bang. The recently concluded
experiments at the CERN SPS and the ongoing experiments at the
Relativistic Heavy Ion Collider at Brookhaven do seem to suggest a very
definite possibility of the confirmation of the quark-hadron phase transition.

It is known for quite some time~\cite{kms} that the transverse momenta
of particles produced in a high multiplicity environment in almost
all models of particle production, roughly 
scale  with the particle rapidity density ($dN/dy$) per unit transverse
area ($\pi R_T^2$),
\begin{equation}
\nu=\frac{1}{\pi R_T^2}\frac{dN}{dy},
\label{nu}
\end{equation}
so that,
\begin{equation}
\langle p_T \rangle \,\sim \, \nu^{1/2}.
\end{equation}
This follows from dimensional arguments as $\nu$ is the only scale
in the problem. This has been noted in minijet models\cite{mini},
 string models\cite{string}, and more recently  suggested
in partonic saturation
descriptions like colour gluon condensate model for heavy ion
 collisions~\cite{cgc}. 

Let us approach this problem from another angle.
Let us assume that in  central nucleus-nucleus collisions at
 relativistic energies
a thermally and chemically equilibrated quark gluon plasma
is formed at an initial time $\tau_0$ at an initial temperature $T_0$.
Once $\nu$ is known and an assumption is made
about the applicability of
Bjorken hydrodynamics~\cite{bj}  we get,
\begin{equation}
\frac{2\pi^4}{45\zeta(3)}\,\frac{1}{\pi R_T^2}\frac{dN}{dy}=4 a
T_0^3\tau_0~~,
\label{T0}
\end{equation}
where $a=42.25\pi^2/90$ for
a plasma of massless quarks (u, d, and s) and gluons and we have 
put the number of flavours as
$\approx$ 2.5 to account for the mass of the strange quarks.
Do the hadronic spectra resulting from the hydrodynamic expansion of the
plasma thus produced show some scaling behaviour?

Assuming that one has $\langle E \rangle = 3T$ for massless particles,
 we may take $\tau_0=1/3T_0$, from considerations of  the uncertainty
relation. Thus 
we see that the initial temperature scales with  $\nu^{1/2}$.
We may also add that it was suggested some-time ago 
that~\cite{sch,scal_phot} the
$S+Au$ and $Pb+Pb$ collisions
at the CERN SPS present an interesting example as they lead to
systems with different transverse sizes but similar initial
temperatures due to the near equality of the scaling variable $\nu$.
(We shall see that the differences for the hadronic spectra for the
two cases are not large.)

In the present work we continue this argument and show that the
average transverse momenta of the produced hadrons,
estimated on the basis of a hydrodynamical expansion with transverse
flow scale with $\nu^{1/2}$. The scaling is 
surprisingly similar to the corresponding behaviour for $p\overline{p}$
data at 1800 GeV, which is sought to be understood 
within a colour gluon condensate model~\cite{cgc}. It may be recalled that
these data were used to provide a suggestion for baryon-flow and quark
gluon plasma~\cite{lm}. A straight forward explanation in
terms of an increased minijet activity which leads to a larger multiplicity
has also been offered for this~\cite{xin}.

The similarity of these behaviours is of interest as
 the hydrodynamics calculations start with a
partonic plasma, which develops into a mixed phase and then into a hadronic
phase. The spectra for particles having different masses are affected
differently due to the celebrated radial flow velocity which develops
during the expansion. More-over several details like the equation of
state, the critical temperature, the freeze-out temperature, and a 
model of hadronization etc. affect the overall evolution of the
system.

This  would seem to suggest that the scaling of the
particle transverse momenta may be `seeded'  in the initial conditions, and
that this behaviour is
broadly retained by the hydrodynamic expansion.  
We insist that radial flow is essential
to obtain this result, as in absence of the radial flow, the 
$\langle p_T \rangle$ is
uniquely determined by the freeze-out temperature and is independent of
multiplicity~\cite{crs} and the size of the system. 
 
The method of solution of the hydrodynamic equations for the
expansion and cooling of the quark-gluon plasma has been discussed
extensively in literature. We only briefly mention the salient features.
As discussed earlier, we assume the plasma to be in a thermally and
chemically equilibrated state of quark gluon plasma, which expands,
cools, undergoes a phase transition to a hot hadronic matter at $T=T_C$
and when the hadronic density is too low to cause further scattering
among the hadrons it undergoes a freeze-out. The plasma is assumed
to undergo a longitudinally boost-invariant and an azimuthally symmetric
transverse expansion.

The initial energy density profile is assumed to follow the so-called
`wounded-nucleon' distribution~\cite{photon}.
We further assume that the phase transition takes place at $T=$ 180 MeV and the
freeze-out takes place at 120 MeV. We
use a hadronic equation of state consisting of {\em {all}} hadrons and
resonances from the particle data table which have a mass less then 2.5
GeV~\cite{crs}. In all the calculations reported here the initial time
$\tau_0$ is taken as 1 fm/$c$. It is well known that as long as
$T_0^3\tau_0$ is kept fixed, a smaller initial time makes only a
marginal difference to the flow.
 The relevant hydrodynamic equations are
solved using the procedure~\cite{hydro} discussed earlier and
spectra for hadrons estimated~\cite{crs} using the Cooper-Fry formulation.

In Fig.1 we show our results for several systems~\cite{sch} having an
identical particle rapidity density per unit transverse area ($\nu$). 
We see that the 
average transverse momenta for pions, kaons, and protons for all the
cases are similar, though they decrease slightly with the
increasing transverse size. This decrease is about 4\% for pions and 
about 7\% for protons.  Even though marginal, this is interesting as 
the radial flow introduces
an additional scale in the problem, the size $R_T$.  At
a time $t$ the region beyond $R_T-c_st$ is affected by the transverse
flow~\cite{munshi}. Thus a smaller system is more strongly affected 
by the flow, even if the initial temperature is similar and thus
build-up~\cite{edward} of the radial flow velocity given by 
\begin{equation}
v_r(t_f) \propto \int_{t_i}^{t_f} \frac{P(t)}{\epsilon(t)}\, dt
\end{equation}
will be affected.  In the above, $P$ is the pressure and $\epsilon$
is the energy density so that the quantity under the integral sign
 gives the speed of
sound which passes though a minimum during the mixed phase.

A demonstration of this effect is seen in Fig.2, where we
have plotted the emission of pions per unit transverse area for the
S+Au (WA80 experiment~\cite{wa80}) and Pb+Pb 
(WA98 experiment~\cite{wa98}) systems
 which have the same $\nu$ but different transverse sizes ($R_T$).
We add that the hydrodynamic treatment used here gives a
perfect description to the hadronic spectra for the $Pb+Pb$ 
system at SPS energy, which correspond to this value of $\nu$~\cite{photon}.
 We note that the inverse slope of the smaller system is 
 marginally larger, as indeed Fig.1 implies. The slight difference in
the radial flow which may develop in systems of varying sizes
can be  magnified by looking at the transverse momenta of
heavier particles. This is seen in Fig.3 where we now plot the  emission of 
protons per unit transverse area for the S+Au (NA35~\cite{na35})
 and Pb+Pb (NA44~\cite{na44}) systems having the same
$\nu$. A much larger freeze-out temperature ( with very different values
 for  pions and protons) would be needed to understand these spectra if
the flow is not included.

In Fig.4 we show the average transverse momenta of pions, kaons, and
protons for central collision of lead nuclei for a range of multiplicity
densities spanning the region likely to be covered at the
RHIC and LHC. It is seen that the average transverse momentum rises with
the square-root of the scaling variable $\nu$.
so that, the pion, kaon, and proton spectra determined from
the hydrodynamics are well described by:
\begin{eqnarray}
\langle p_T \rangle&=&0.326+0.052\sqrt{\nu} ~;~\pi~,\nonumber\\
\langle p_T \rangle&=&0.423+0.087\sqrt{\nu} ~;~K~,\nonumber\\
\langle p_T \rangle&=&0.530+0.137\sqrt{\nu} ~;~p~.
\label{pt}
\end{eqnarray}
We also find (see Fig.5) that the
the hadronic-spectra scale  as
\begin{equation}
\frac{dN_h}{d^2p_Tdy}=\frac{\alpha}{\langle p_T \rangle ^2}
 F\left(\frac{p_T}{\langle p_T \rangle }\right).
\label{scale}
\end{equation}

The preliminary data for the transverse momenta are now available from
the PHENIX experiment ~\cite{julia} at $\sqrt{s_{NN}}$ 130 GeV. We see
from the Fig.6 that these data conform to the hydrodynamic
behaviour seen in this work.

In Fig.7 we compare the hydrodynamics results for $Pb+Pb$ system with
 the experimental results for the $p\overline{p}$ collision at
 1800 GeV~\cite{exp}. Their agreement points to a universal scaling
 behaviour fot the transverse momentum distributions.
Considering that several models of particle production relate the
increase in $\langle p_T \rangle>$ with the multiplicity to the
 initial conditions,
we have plotted the $\langle  p_T \rangle $ of the hadrons against the
 initial temperature
obtained and used in the hydrodynamic calculations here and we see (Fig.8)
 that the increase is directly related to initial temperature. 

It has been suggested~\cite{bl} that hydrodynamic flow at the end of the
three dimensional evolution of the plasma is uniquely determined by
a dimensionless parameter $\omega$:
\begin{equation}
\omega=s_0\tau_0/s_cR_T
\end{equation}
where $s_0$ is initial entropy density, $\tau_0$ is initial time, $s_c$
is the entropy density at the beginning of the phase-transition, and $R_T$
is the transverse size. Recalling that $s \sim T^3$, one can easily
prove that $\omega=\tau_c/R_T$, where $\tau_c$ is the proper time
when the phase transition begins, provided the QGP cools according
to a boost-invariant longitudinal expansion. Obviously if $\omega$ is
small, transverse flow effects should be small, as the QGP phase is over
well before $R_T/c_s$, when the rare-faction wave covers the distance $R_T$
and agitates the entire fluid. Using Eq.(\ref{T0}), we can also
see that $\omega\sim (dN/dy)/A$. We now investigate this scaling
in terms of a variable 
\begin{equation}
\mu=\frac{1}{A}\,\frac{dN}{dy}
\end{equation}
in the spirit of a similar study by Ruuskanen~\cite{rus_pol}.

In Fig.9 we have plotted the average transverse momenta of hadrons
obtained for $S+S$, $Zr+Zr$ and $Pb+Pb$ systems for a large range of 
values of the scaling variable $\mu$. We see that the professed 
scaling is only approximately satisfied, and becomes worse with
increasing mass of the hadron at large $\mu$.  This could be
due to the fact that for such
values the development of the flow after the conversion to hadrons
starts, can not be ignored. We also note that the this scaling is
not universal as the the corresponding results for the $p\overline{p}$
data show a very different behaviour.

Before concluding, let us discuss the uncertainties in our estimates.
The hydrodynamics
 calculations have several inputs which can slightly alter the extent of the
deviations of the scaling behaviour observered here.
 We have used a freeze-out temperature of 120 MeV. Decreasing
$T_F$ increases the $\langle p_T \rangle $ slightly, the largest increase  being
for protons. For $\nu$=5, decreasing $T_F$ to 100 MeV increases the
$\langle p_T \rangle$ for pions from 0.44 GeV/$c$ to 0.46 GeV/$c$,  while for
protons the corresponding increase is from 0.83 GeV/$c$ to 1.02 GeV/$c$.
For $\nu=50$, the $\langle p_T \rangle$ for pions increases from 0.70 GeV/$c$ to
0.72 GeV/$c$ when the freeze-out temperature is decreased to 100 MeV,
while the increase for protons is from 1.5 GeV/$c$ to 1.7 GeV/$c$.
The critical temperature used affects the $\langle p_T \rangle$ too, 
in a very interesting
manner. Increasing the $T_C$ increases the degrees of freedom in the
hadronic matter (as higher resonances get excited) and the phase-transition
is completed quickly. Thus the duration of the 
low-pressure gradient associated with
the mixed phase is reduced and $\langle p_T \rangle$ increases slightly,
 and the opposite
happens when a lower value of $T_C$ is used. We have verified that the
change in $\langle p_T \rangle$ for protons is less than 10\% when
 $T_C$ is  changed by
$\pm$ 20 MeV. For pions the results are much less altered. It is easy
to argue that if a much simpler equation of state for hadrons is
used (say having only pions) then the mixed phase will live much 
longer and $\langle p_T \rangle$ will decrease.

It is also useful to remember that 
the $p_T$ broadening and the so-called jet-quenching should
also affect results discussed in Fig.3, e.g. It is hard to decide
the relative worth of hydrodynamic and such treatments
till precise data at higher energies and upto large $p_T$ are available.

The discussion so far has involved central collisions, as it is not easy
to extend the hydrodynamic treatment to non-central (non-zero impact
parameter) collisions. A rough estimate could still be made if we
ignore the azimuthal asymmetry of the region of overlap, approximating
it to a circle having a radius $R\approx 1.2 (N_{\mbox{\rm{part}}}/2)^{1/3}$,
where $N_{\mbox{\rm{part}}}$ is the number of participants.
This would then help assign a value of $\nu$ even to non-central
collisions, as has been done in Ref.~\cite{cgc}. Recall that the 
NA49 data for $Pb+Pb$ collisions treated in this manner were found to
similar to the E735 data for $p\overline{p}$ in the above work 
bolstering the arguments for a universal behaviour. We eagerly await
the data from RHIC experiments to verify this aspect.

The similarity of the behaviour observed here for high multiplicity
$p\overline{p}$ and $AA$ results raises one question, which is
applicable to all the studies involving hydrodynamics; viz., are
the conditions appropriate for the applicability of hydrodynamics 
realized in these situations? The similarity of the behaviour for the
high multiplicity events in $p\overline{p}$ with those of $AA$ does
suggest a plausible argument. It indicates that the 
the initial states of these systems are perhaps consisting of partons
and the multiple collisions indeed drive the system to an equilibrium
quickly. It is known from several studies that high multiplicity
events in hadron-hadron collisions are also rich in multiple 
partonic collisions~\cite{xin} and provide a basis for the
parton cascade model~\cite{klaus}.

In brief, we see that hadronic spectra generated from boost-invariant
hydrodynamics with transverse expansion show a scaling behaviour
in the particle rapidity density per unit transverse area. For smaller
values of the scaling variable $\nu$, the results are quite close
to values inferred from ISR data, which are sought to be understood
in terms of a universal behaviour given by the
colour gluon condensate model. 

\section{Acknowledgements}*
 The author is grateful to  Larry McLerran for his searching
queries and useful discussions. Valuable comments from  
Berndt M\"uller, J. -Y. Ollitrault, 
J. Schaffner-Bielich, and Xin Nian Wang are also gratefully acknowledged.
He is also grateful to Prof. J. Velkovska for sending him the
numerical values of the preliminary PHENIX data.

\newpage

\begin{figure}
\epsfxsize=3.25in
\epsfbox{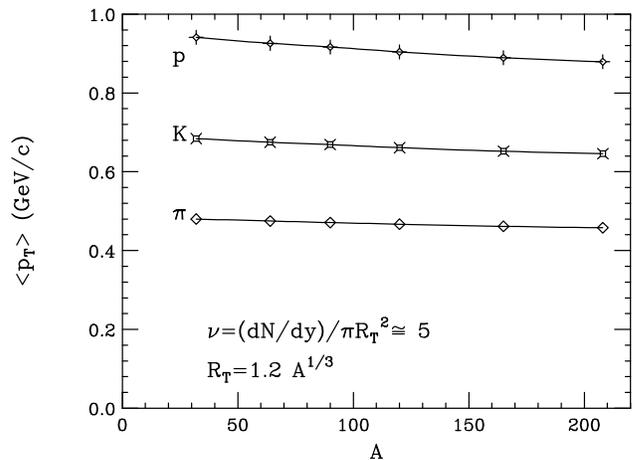}
\vskip 0.2in
\caption{ 
A model calculation of the
average transverse momentum of hadrons for several systems which
have identical particle rapidity density per unit transverse area.
} 
\end{figure}

\begin{figure}
\epsfxsize=3.25in
\epsfbox{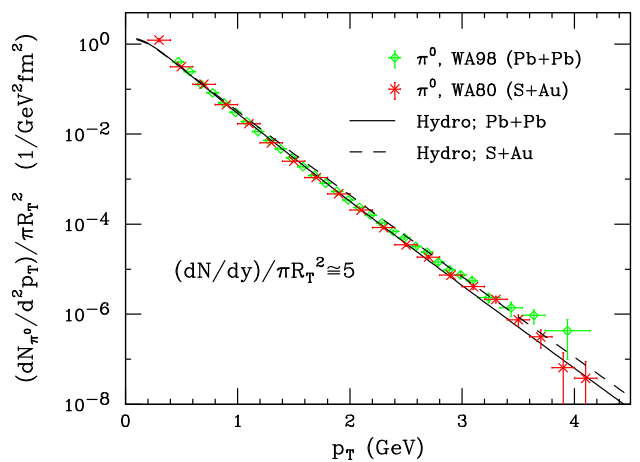}
\vskip 0.2in
\caption{ 
Emission of pions per unit transverse area for  S+Au 
(WA80\protect\cite{wa80}) and Pb+Pb (WA98\protect\cite{wa98})
systems, which  have the
same multiplicity per unit transverse area but different transverse
sizes.
} 
\end{figure}

\newpage

\begin{figure}
\epsfxsize=3.25in
\epsfbox{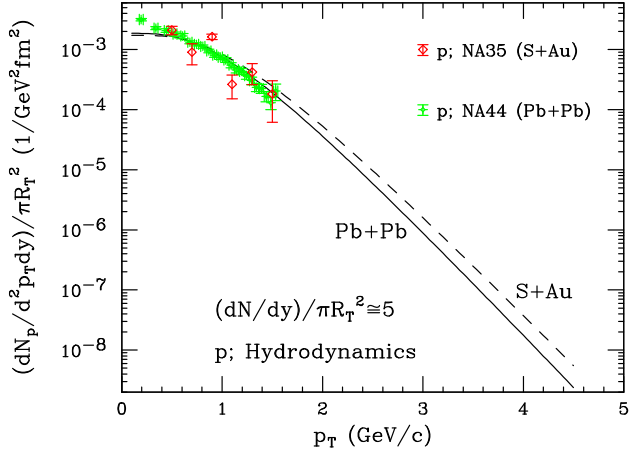}
\vskip 0.2in
\caption{Ehe emission of
of protons per unit transverse area for
$S+Au$ and $Pb+Pb$ having same $\nu$.
} 
\end{figure}

\begin{figure}
\epsfxsize=3.25in
\epsfbox{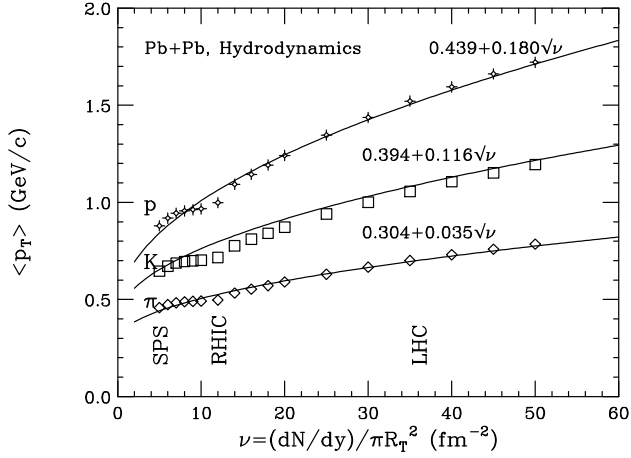}
\vskip 0.2in
\caption{
A model calculation of the
average transverse momentum of hadrons for central collision of $Pb$ nuclei
as a function of the particle rapidity density per unit transverse area.
The symbols represent the hydrodynamics results while the solid curves give
the fits described in the text (Eq.\protect\ref{pt}).
}
\end{figure}

\newpage

\begin{figure}
\epsfxsize=3.25in
\epsfbox{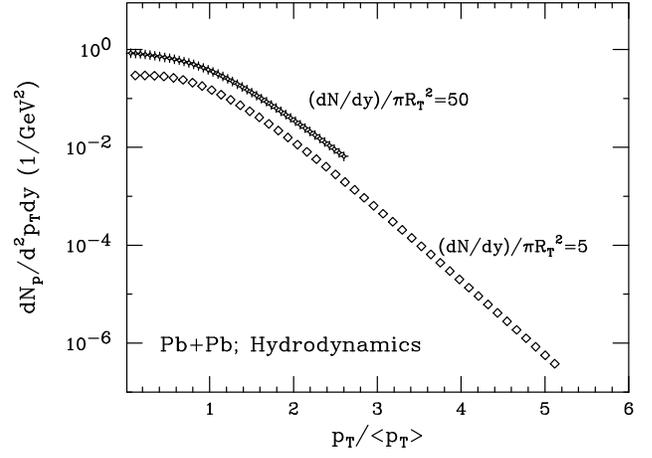}
\vskip 0.2in
\caption{
Scaling of hadronic spectra obtained from hydrodynamics
(Eq.\protect\ref{scale}). The two curves differ by a factor which is 
equal to the ratio of the square of the average transverse momenta (see text).
}
\end{figure}

\begin{figure}
\epsfxsize=3.25in
\epsfbox{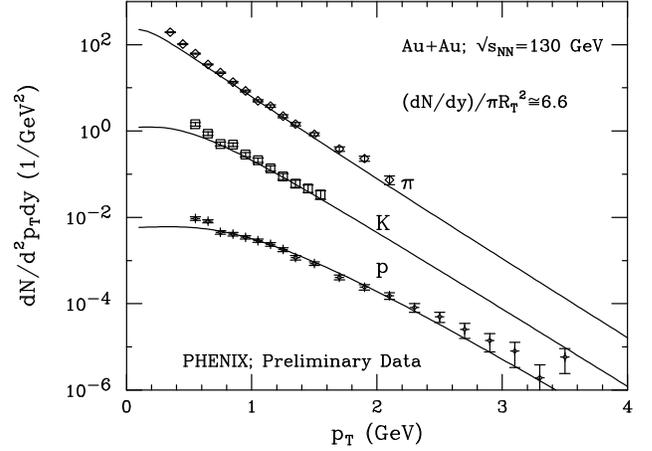}
\vskip 0.2in
\caption{
Hadronic spectra from PHENIX experiment \protect\cite{julia}.
}
\end{figure}

\newpage

\begin{figure}
\epsfxsize=3.25in
\epsfbox{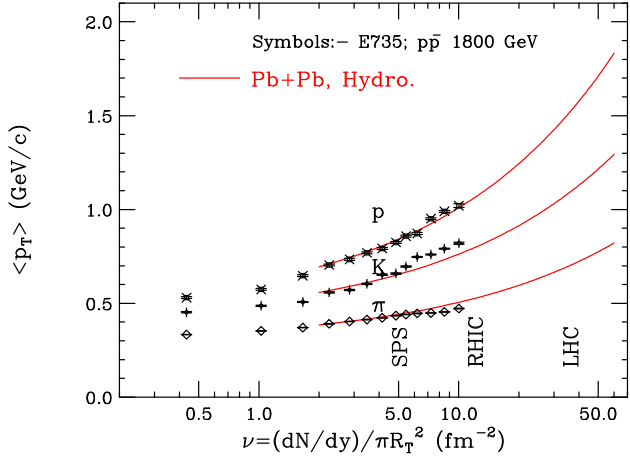}
\vskip 0.2in
\caption{
A comparison of scaling suggested by hydrodynamics (solid curves)
and the corresponding  data for \protect$p\overline{p}$ scattering at 1800 GeV
\protect\cite{exp}.}
\end{figure}

\begin{figure}
\epsfxsize=3.25in
\epsfbox{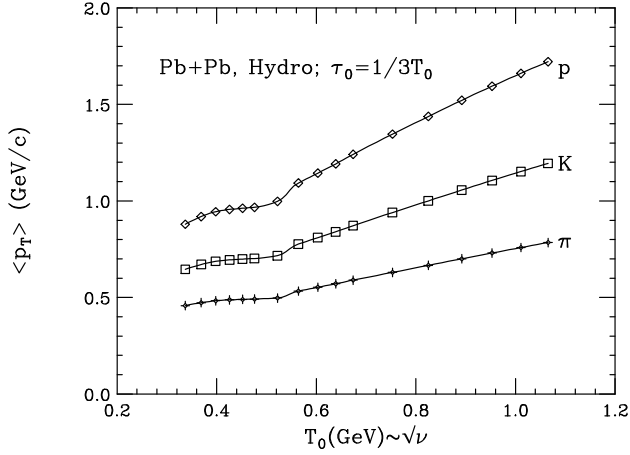}
\vskip 0.2in
\caption{
The average transverse momenta attained as a function of the initial
temperature. 
}
\end{figure}
\newpage
\begin{figure}
\epsfxsize=3.25in
\epsfbox{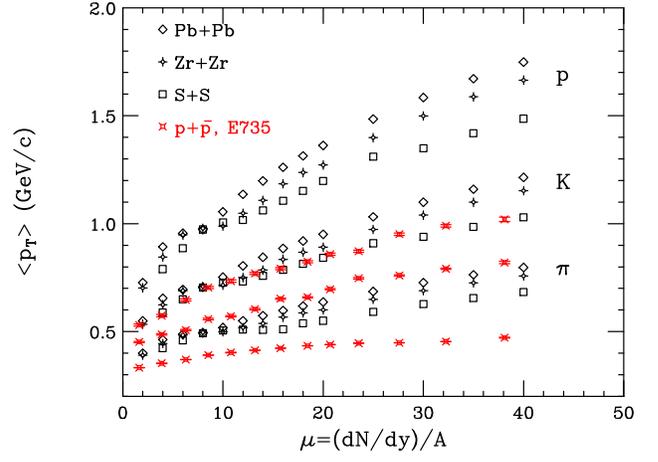}
\vskip 0.2in
\caption{
The average transverse momenta for pions, kaons, and protons for
central collisions of $S+S$, $Zr+Zr$, and $Pb+Pb$ systems as a
function of multiplicity per nucleon. The corresponding values 
for the $p\overline{p}$ data \protect\cite{exp} are also given.}
\end{figure}

\end{document}